\documentclass[10pt,conference]{IEEEtran} 
\usepackage[utf8]{inputenc}
\usepackage{cite}
\usepackage{amsmath,amssymb,amsfonts}
\usepackage{amsthm}

\usepackage{algorithmic}
\usepackage{graphicx} 
\usepackage{booktabs}
\usepackage{textcomp}
\usepackage{subcaption}
\usepackage{tcolorbox}
\usepackage{amsmath}
\usepackage{amssymb}
\usepackage{enumerate}
\usepackage{enumitem}
\usepackage{alltt}
\usepackage{url}
\usepackage{multirow}
\usepackage{balance}

\theoremstyle{definition}
\newtheorem{defn}{Definition}

\usepackage{listings}
\lstdefinelanguage{diff}{
  language=Java,
  numbers=none,
  commentstyle=\color{gray}, 
  morecomment=[f][\color{blue}]{@@},     
  morecomment=[f][\color{red!60!black}]-,         
  morecomment=[f][\color{green!60!black}]+,       
  morecomment=[f][\color{magenta}]{---}, 
  morecomment=[f][\color{magenta}]{+++},
}
\usepackage{stackengine}
\usepackage{boxhandler}
\usepackage{enumitem}
\usepackage{listings}             
\usepackage{xcolor}

\makeatletter
\let\old@lstKV@SwitchCases\lstKV@SwitchCases
\def\lstKV@SwitchCases#1#2#3{}
\makeatother
\usepackage{lstlinebgrd}
\makeatletter
\let\lstKV@SwitchCases\old@lstKV@SwitchCases

\lst@Key{numbers}{none}{%
    \def\lst@PlaceNumber{\lst@linebgrd}%
    \lstKV@SwitchCases{#1}%
    {none:\\%
     left:\def\lst@PlaceNumber{\llap{\normalfont
                \lst@numberstyle{\thelstnumber}\kern\lst@numbersep}\lst@linebgrd}\\%
     right:\def\lst@PlaceNumber{\rlap{\normalfont
                \kern\linewidth \kern\lst@numbersep
                \lst@numberstyle{\thelstnumber}}\lst@linebgrd}%
    }{\PackageError{Listings}{Numbers #1 unknown}\@ehc}}
\makeatother

\definecolor{backcolour}{rgb}{0.95,0.95,0.92}

\lstdefinestyle{codestyle}{
    commentstyle=\color{green},
    keywordstyle=\color{blue},
    numberstyle=\scriptsize\color{gray},
    stringstyle=\color{purple},
    basicstyle=\ttfamily\scriptsize,
    breakatwhitespace=false,         
    breaklines=true,                 
    captionpos=b,                    
    keepspaces=true,                 
    numbers=left,                    
    numbersep=5pt,                  
    showspaces=false,                
    showstringspaces=false,
    showtabs=false,                  
    tabsize=2,
    frame=none
}

\newcommand{\DeltaSpec}{\textsc{DeltaSpec}}

\newcommand{\CodeIn}[1]{{\small \texttt{#1}}}

\begin{document}
\title{Specification Inference for Evolving Systems}

\author{
\IEEEauthorblockN{Renzo Degiovanni\IEEEauthorrefmark{1}, Facundo Molina\IEEEauthorrefmark{2}, Agustin Nolasco\IEEEauthorrefmark{3}, Nazareno Aguirre\IEEEauthorrefmark{3} and Mike Papadakis\IEEEauthorrefmark{1}}
\IEEEauthorblockA{\IEEEauthorrefmark{1}University of Luxembourg, Luxembourg}
\IEEEauthorblockA{\IEEEauthorrefmark{2}IMDEA Software Institute, Spain}
\IEEEauthorblockA{\IEEEauthorrefmark{3}University of Río Cuarto, Argentina}
}

\maketitle

\begin{abstract}
Regression tests are often used to ensure that software behaves as intended, as software evolution takes place. However, tests in general and regression tests in particular are inherently partial as behavioural descriptions, and thus regression tests may fail in clearly capturing how software behaviour is altered by code changes. In this paper, we propose an assertion-based approach to capture software evolution, through the notion of \emph{commit-relevant specification}. A commit-relevant specification summarises the program properties that have changed as a consequence of a \emph{commit} (understood as a specific software modification), via two sets of assertions, the \emph{delta-added assertions}, properties that did not hold in the pre-commit version but hold on the post-commit, and the  \emph{delta-removed assertions}, those that were valid in the pre-commit, but no longer hold after the code change. We also present \DeltaSpec{}, an approach that combines test generation and dynamic specification inference to automatically compute commit-relevant specifications from  given commits. We evaluate \DeltaSpec{} on two datasets that include a total of 57 commits (63 classes and 797 methods). We show that commit-relevant assertions can precisely describe the semantic deltas of code changes, providing a useful mechanism for validating the behavioural evolution of software. We also show that \DeltaSpec{} can infer 88\% of the manually written commit-relevant assertions expressible in the language supported by the tool. Moreover, our experiments demonstrate that \DeltaSpec{}'s inferred assertions are effective to detect regression faults. 
More precisely, we show that commit-relevant assertions can detect, on average, 78.3\% of the artificially seeded faults that interact with the code changes. We also show that assertions in the delta are 58.3\% more effective in detecting commit-relevant mutants than assertions outside the delta, and that it takes on average 169\% fewer assertions when these are commit-relevant, compared to using general valid assertions, to achieve a same commit-relevant mutation score. 
\end{abstract}

\section{Introduction}
\label{sec:introduction}



Software testing is one of the main techniques typically used to ensure software evolution progresses as expected \cite{PezzeYoung2007}. 
Indeed, testing is used to verify that the program behaviour
is preserved after changes 
and to  
exercise new/changed functionality in the new version of the software, via the so-called regression tests \cite{YooH12, RothermelH94}. This means that the employed test cases and their corresponding test assertions, reflect the behavioural evolution of the software, in the \emph{scenario-specific} manner inherent to testing. As a consequence, in the same way that tests may be too partial as a description of software behaviour \cite{DBLP:books/sp/Meyer14}, regression tests can also be too scenario-specific, as a description of the evolution of software behaviour. 

More generally, software behaviour descriptions can come in different forms, e.g., as natural language documentation (software documents, code comments), as well as in the form of \emph{formal program specifications}, i.e., (executable) expressions that capture the expected behaviour of programs at specific program locations \cite{Meyer1997,Schiller-TypeContracts}. We are interested in this kind of specifications, but not as general descriptions of software, but as descriptions of the software \emph{evolution}. More precisely, in the same way that version control software can highlight syntactic modifications associated with a specific software change, what is typically referred to as the \emph{delta} of a \emph{code commit} \cite{DBLP:books/daglib/0014807}, we are interested in emphasising the \emph{semantic delta} of the commit, i.e., what software properties have changed as a consequence of a software modification. Consider for instance the following code change extracted from the Commons Math repository (corresponding to commit \CodeIn{ce18534}\footnote{https://github.com/apache/commons-math/commit/ce18534}):

\begin{lstlisting}[language=diff]
  public double getLInfNorm() {
    double max = 0;
    for (double a : data) {
-     max += Math.max(max, Math.abs(a));
+     max = Math.max(max, Math.abs(a));
    }
    return max;
  }
\end{lstlisting}

\noindent
Notice how the change is remarked indicating the \emph{previous} code, and the \emph{updated} code, for more convenient comparison. If we consider the effect of this syntactic change in the behaviour of method \CodeIn{getLInfNorm()}, in particular in relation to the method's postcondition, a possible specification of the \emph{delta}, that we call a \emph{commit-relevant specification}, would be the following (ignoring, for a moment, overflow issues):

\begin{lstlisting}[language=diff,escapeinside={(*}{*)}]
- result >= (*{\color{red!60!black}$\sum_{i=0}^{N-1} abs(data[i])$}*)

+ result = (*{\color{green!60!black}$Max_{i=0}^{N-1}\, abs(data[i])$}*)
\end{lstlisting}

\noindent
This commit-relevant specification states that it is not longer the case that the result of the method is always greater or equal than the summation of the absolute values of the array elements (the property held before the commit, but is violated in the updated code); and now the result of the method is the maximum among the absolute values of the array elements (it did not hold in the previous version, it emerges as a new property of the updated code). It is worth mentioning that the unit tests documenting this commit are less informative: they simply state that method \CodeIn{getLInfNorm()} should return 6 when called on array $[-4, 0, 3, 1, -6, 3]$ (cf. \CodeIn{ArrayRealVectorTest.java} and \CodeIn{SparseRealVectorTest.java} in commit \CodeIn{ce18534}).


The rationale of commit-relevant specifications is that they \emph{explain} the specific semantic modifications to the source code, generated by a commit. Their inference can be used for manual inspection, as a way of validating a change, as well as for automated analyses, e.g., for testing or bug finding. Since inferring specifications in general, and commit-relevant specifications in particular, can be challenging and time consuming, we introduce \DeltaSpec{}, an approach based on test generation and dynamic specification inference, that automatically infers commit-relevant specifications. \DeltaSpec{} takes as input the pre-commit version $P_{pre}$ and the post-commit version $P_{post}$ of a program $P$, and returns a commit-relevant specification composed of two sets of commit-relevant assertions:
\begin{description}
\item[delta-added assertions:] assertions that are valid in the post-commit version but were invalid in the pre-commit version; and
\item[delta-removed assertions:] assertions that are invalid in the post-commit version but were valid in the pre-commit version.    
\end{description}
\DeltaSpec{} works by first using \emph{test generation} to produce test suites for both the $P_{pre}$ and  $P_{post}$ versions of $P$. It then uses a \emph{specification inference} tool to infer program specifications $S_{pre}$ and $S_{post}$, capturing the behaviours of $P_{pre}$ and $P_{post}$, according to their corresponding test suites. Finally, the commit-relevant assertions for this code modification is obtained by processing $S_{pre}$ and $S_{post}$ to compute a ``diff''. 


We implement \DeltaSpec{} and evaluate our proposal on two aspects. First, we study the adequacy of commit-relevant specifications for capturing and explaining the semantic delta of code modifications, and the effectiveness of \DeltaSpec{} to infer expected commit-relevant assertions. We analyze 18 commits from the Apache Commons~\cite{apache_commons} family, and show that  56 manually produced commit-relevant assertions effectively capture the corresponding code changes; moreover, \DeltaSpec{} is able to reproduce 88\% of the manually written commit-relevant assertions expressible in its supported specification language. Second, we study a concrete use case for commit-relevant assertions, their usage in the context of \emph{commit-aware mutation testing}~\cite{ma2020commit} to assess test suite adequacy. We take a total of 39 commits from a publicly available dataset~\cite{DBLP:journals/tosem/OjdanicSDPL2022}, again from Apache Commons projects, and show that commit-relevant assertions produced by \DeltaSpec{} can detect an average of 78.3\% commit-relevant mutants, i.e., artificial faults related to the commit. Moreover, we observe that if commit-relevant assertions are selected instead of non-commit-relevant valid assertions for the post-commit, the commit-relevant mutation score is increased by 58.3\%, on average. Similarly, it takes on average 169\% fewer assertions when these are taken from the set of commit-relevant assertions, compared to using general valid assertions of the post-commit, to achieve a same commit-relevant mutation score.

In summary, our paper makes the following contributions:
\begin{enumerate}
\item We introduce the notion of \emph{commit-relevant specifications} to explain the semantic delta generated by a specific commit change.
\item We present \DeltaSpec{}, the first automatic approach to automatically infer commit-relevant specifications. We show that \DeltaSpec{} is very effective in reproducing commit-relevant specifications written by developers (88\% of those expressible in the language supported by the tool).
\item We show that commit-relevant assertions are effective for testing commit changes. On average, their use can detect 78.3\% of commit-relevant mutants, 58.3\% more mutants than those detected by using valid assertions of the post-commit, outside the delta. Also, it takes on average 169\% fewer assertions when these are commit-relevant, compared to using general valid assertions, to achieve a same commit-relevant mutation score.
\end{enumerate}

\section{Preliminaries}

\subsection{Testing Evolving Systems}
\label{sec:CI-testing}
Mutation testing~\cite{PapadakisK00TH19} is a test adequacy criterion where test requirements are represented by \emph{mutants}, i.e., artificially seeded faults obtained from slight syntactic modifications to an original program (e.g., \CodeIn{x > 0} is mutated to \CodeIn{x < 0}). Mutants are used to assess the effectiveness and thoroughness of a test suite, by measuring how many of these artificial faults the suite is able to detect. A mutant is said \emph{killed} if there exists a test case that is capable of producing distinguishable observable outputs between the mutant and the original program. Otherwise, the mutant is called \emph{live}, or is said to have \emph{survived}. Some mutants cannot be killed as they are functionally \emph{equivalent} to the original program. The \emph{mutation score} (MS) is the measure computed as the ratio between killed mutants over the total number of generated mutants. 

Software is under constant evolution. Developers update their software as part of debugging, code improvements, and adapting code to changing requirements, among other tasks. Moreover, modern software development promotes incremental implementation of software, with code bases being continuously updated with frequent code contributions supported by Continuous Integration (CI) and related practices~\cite{handsOnDevOps}. In this setting of constantly evolving software, it is important to test the program modifications to avoid introducing faults. To do so, developers typically perform \emph{regression testing}, the process of writing (additional) test cases that exercise the changes, stress their dependencies, and check that the program changes behave as intended~\cite{YooH12, HenardPHJT16, RothermelH94}. Since applying traditional mutation testing to the entire code base is typically infeasible in CI contexts, \emph{commit-aware mutation testing} aims at focusing the mutation testing analysis on the program behaviours affected by code changes~\cite{petrovic2018, ma2020commit, ma2021mudelta, DBLP:journals/ese/OjdanicMLCVP22, DBLP:journals/tosem/OjdanicSDPL2022}. Commit-aware mutation testing considers \emph{commit-relevant mutants}, i.e., those whose behaviour is affected by the corresponding code change. Indeed, Ma et al.~\cite{ma2020commit} define a  mutant to be relevant with respect to a commit if its execution on the pre-commit is different from its execution in the post-commit.  

Commit-relevant mutants can be used as the test requirements to guide the testing process and assess the thoroughness of a test suite with respect to the committed change. The \emph{commit-relevant mutation score} (rMS) is computed as the ratio between the number of killed commit-relevant mutants and the total number of commit-relevant mutants generated. 


We will assess commit-relevant assertions in the context of commit-relevant mutation testing, by analysing if these assertions can effectively identify commit-relevant mutants. 

\subsection{Specification Inference}
\label{subsec:spec-inference}

Specification inference refers to the task of automatically producing a formal description of the software behaviour from existing software artifacts, such as documentation, source code, code comments, etc. The inferred specifications can be used as an \emph{oracle}~\cite{Barr+2015} that distinguishes correct/intended program behaviour from incorrect/unintended program behaviour. In recent years, the specification inference problem has gained an increasing attention from the software engineering community, leading to the proposal of various tools and techniques for automated specification inference (some examples are Daikon~\cite{DBLP:journals/scp/ErnstPGMPTX07}, Jdoctor~\cite{Blassi+2018}, GAssert~\cite{gassert2020}, EvoSpex~\cite{Molina+2021}, and SpecFuzzer~\cite{Molina+2022}).

More formally, a program specification for a particular program point is a set of assertions that are valid in \emph{every} execution of the program, at the corresponding program point. These assertions are also referred to as \emph{program invariants}~\cite{DBLP:journals/scp/ErnstPGMPTX07}. Let us now introduce some notation and definitions regarding the specification language and program assertions that we use in the paper.

\begin{defn}[Program Assertions]
\label{def:program-assertions}
Let $P$ be a program, and ${\cal L}$ a formal logical language. Given a program point $p$ of $P$, a sentence $\varphi \in {\cal L}$ is said to be an \emph{assertion} for $p$ if and only if for every execution $t$ of $P$ and state $s_P$ of $t$ at location $p$, $\varphi$ can be evaluated at $s_P$. The set of assertions for program $P$ at location $p$ according to ${\cal L}$ is denoted by ${\cal T}_{P, p}^{\cal L}$.
\end{defn}

\begin{defn}[Program Specification]
\label{def:program-specification}
Let $P$ be a program, and ${\cal L}$ a formal language. Given a reachable program point $p$ of $P$, the specification ${\cal S}_{P, p}^{\cal L}$ for program $P$ at location $p$ according to ${\cal L}$ is defined as the set of assertions for program $P$ at location $p$ such that, for every execution $t$ of $P$ and state $s_P$ of $t$ at location $p$, $\varphi$ holds (i.e., evaluates to true) at $s_P$. 
\end{defn}

We introduce the notion of commit-relevant specifications, and a technique, \DeltaSpec{}, to infer these specifications.  \DeltaSpec{} relies on existing specification inference tools. More precisely, our technique relies on SpecFuzzer~\cite{Molina+2022}, a tool that extends and uses Daikon~\cite{DBLP:journals/scp/ErnstPGMPTX07} for specification inference. SpecFuzzer uses grammar-based fuzzing to automatically generate candidate assertions for a target program, and then uses Daikon~\cite{DBLP:journals/scp/ErnstPGMPTX07} to determine which of those assertions are consistent with the behaviours exhibited by a provided test suite. The main reason for choosing SpecFuzzer is that it can be adjusted to different assertion languages by tailoring a grammar, as opposed to other approaches that may require making more complex changes in the respective tools. 

\section{Motivating Example}
\label{sec:motivation}

Figure~\ref{example-math-d4j-206-change} illustrates a fragment of commit \CodeIn{bfe4623}\footnote{https://github.com/apache/commons-math/commit/bfe4623} from the Apache Commons Math Library\footnote{https://commons.apache.org/proper/commons-math/}, which we use to motivate our work. To understand the rationale of this commit, let us first introduce the issue being solved. The commit change is affecting the \CodeIn{Sum} class, which allows one to compute a summation of a list of doubles, through successive invocations of method \CodeIn{increment}. Figure~\ref{example-math-d4j-206-usage} completes our illustration, featuring a snippet of class \CodeIn{OneWayAnovaImpl}, which uses class \CodeIn{Sum}: it generates an object of class \CodeIn{Sum}, and then iterates over all the double values in the \CodeIn{data} array using \CodeIn{increment}.

\begin{figure}[t]
\vspace{-0.5em}
\centering
\begin{subfigure}{.4\textwidth}
\begin{lstlisting}[language=diff]
  public Sum() {
    n = 0;
-   value = Double.NaN;
+   value = 0;
  }
  /**
  @@ -70,11 +71,7 @@ public Sum(Sum original) {
  */
  public void increment(final double d) {
-   if (n == 0) {
-     value = d;
-   } else {
-     value += d;
-   }
+   value += d;
    n++;
  }
  public double getResult() {
    return value;
  }
\end{lstlisting}
\end{subfigure}
\vspace{-1.0em}
\caption{Bug-fixing commit, from commons-math repository.}
\label{example-math-d4j-206-change}
\vspace{-1.0em}
\end{figure}


\begin{figure}[htp!]
\vspace{-0.5em}
\centering
\begin{lstlisting}[language=diff,numbers=none]
  private AnovaStats anovaStats(Collection<double[]> categoryData) {
    ...
    Sum sum = new Sum();
    SumOfSquares sumsq = new SumOfSquares();
    int num = 0;
    for (int i = 0; i < data.length; i++) {
      double val = data[i];
      num++;
      sum.increment(val);
      sumsq.increment(val);
      ...
    }
    ...
    double ss = sumsq.getResult() - sum.getResult() 
                    * sum.getResult() / num;
    ...
  }
\end{lstlisting}
\vspace{-1.0em}
\caption{Class \CodeIn{OneWayAnovaImpl} using  the \CodeIn{Sum} class.}
\label{example-math-d4j-206-usage}
\vspace{-0.8em}
\end{figure}

The pre-commit version of class \CodeIn{Sum} has a problem: when invoked with an empty \CodeIn{data} array, the attribute \CodeIn{value} is initialized with value \CodeIn{Double.NaN}. This situation, as reported in a corresponding issue\footnote{https://issues.apache.org/jira/browse/MATH-373}, is inconsistent with the mathematical notion of the summation: the summation of an empty array should return \texttt{0} instead of \CodeIn{Double.NaN}. The commit shown in Figure~\ref{example-math-d4j-206-change} fixes this issue by performing two code changes. First, it changes the constructor to initialize attribute \CodeIn{value} with \CodeIn{0}. Second, it updates method \CodeIn{increment}: since now attribute \CodeIn{value} is initialized with \CodeIn{0}, there is no need to distinguish the first invocation to the method from the others, and directly accumulates \CodeIn{d} into attribute \CodeIn{value}. Finally, the increment of attribute \CodeIn{n}, which counts the number of invocations to the method, remains intact. 


In a code maintenance scenario, developers modifying a code base may be interested in validating their changes. As described in section~\ref{subsec:spec-inference}, assertion inference techniques can produce formal descriptions of the program behaviour, that can then be used to understand the current behaviour. Thus, developers can benefit from inferred program assertions to analyse how these assertions evolve/change throughout the evolution of the code base. For instance, Figure~\ref{fig:pre-commit-spec} shows assertions for the pre-commit version, that a developer can obtain by using a specification inference tool such as SpecFuzzer~\cite{Molina+2022}. It is straightforward to see that in this pre-commit version, if method \CodeIn{increment} was not invoked (i.e., \CodeIn{n == 0}), attribute \CodeIn{value} will have value \CodeIn{Double.NaN}.

\begin{figure}[t]
\vspace{-0.5em}
\centering
\begin{lstlisting}[language=diff,numbers=none]
 // Sum class invariant
 (n == 0) implies (value == Double.NaN)
  n >= 0
  
 // increment method
 (n == 0) implies value = d
 (n > 0) implies value = \old(value) + d
  n = \old(n) + 1
\end{lstlisting}
\vspace{-1.0em}
\caption{Specification for the pre-commit version of class \CodeIn{Sum}.}
\label{fig:pre-commit-spec}
\vspace{-0.5em}
\end{figure}

The developer may again infer specifications on the post-commit version of the program, to understand the new behaviour of the \CodeIn{Sum} class. But since this is a second ``fresh'' inference execution, and inference tools typically involve some random or non-deterministic behaviour, extracting what ``has changed''. by simple comparison between the two sets of inferred assertions may be far from trivial.

In order to understand what software properties have changed as a consequence of code changes, is that we introduce the notion of \emph{commit-relevant specification}. 
Intuitively, commit-relevant specifications aim at capturing the \emph{semantic delta} between the pre-commit and post-commit versions in terms of a set of assertions whose validity (invalidity) is affected by the \emph{syntactic delta}, i.e., the committed code changes.  
In this context, a \emph{delta-added assertion} captures a software property that emerges as a consequence of the change, i.e., the assertion does not hold in the pre-commit version but holds in the post-commit version. 
While a \emph{delta-removed assertion} captures properties that are no longer valid as a consequence of the change, i.e., the assertion held in the pre-commit version but does not hold in the post-commit version. 

Considering the commit in Figure~\ref{example-math-d4j-206-change}, since the change is only affecting the \CodeIn{value} attribute, a commit-relevant specification for such a change should only reflect the changes over that attribute. Figure~\ref{example-math-d4j-206-manual-delta-spec} illustrates a (manually written) candidate  \emph{commit-relevant specification} for this commit. In red we indicate the delta-removed assertions, and in green the delta-added assertions. It is straightforward to see how these assertions relate to the changes made in the code. In terms of the \CodeIn{Sum} class, we express that when \CodeIn{n == 0} (i.e., no increment has been performed) the value of \CodeIn{value} is now \CodeIn{0} instead of \CodeIn{Double.NaN}. Regarding the change in the \CodeIn{increment} method we express the removal of the conditional update and the emergent property indicating how the \CodeIn{value} attribute is updated. Notice that we also express properties over \CodeIn{n}, that remain valid across the two commit versions. 

\begin{figure}[t]
\vspace{-0.5em}
\centering
\begin{lstlisting}[language=diff,numbers=none]
  // Sum class invariant
- (n == 0) implies (value == Double.NaN)
+ (n == 0) implies (value == 0)
  n >= 0
  
  // increment method
- (n == 0) implies value = d
- (n > 0) implies value = \old(value) + d
+ value = \old(value) + d
  n = \old(n) + 1
\end{lstlisting}
\vspace{-1.0em}
\caption{Manually written commit-relevant specification for the commit in Figure~\ref{example-math-d4j-206-change}.}
\label{example-math-d4j-206-manual-delta-spec}
\vspace{-0.5em}
\end{figure}


To provide developers with a mechanism that facilitates the understanding and validation of semantic deltas, we develop \DeltaSpec{}, the first fully-automated technique to infer commit-relevant specifications. \DeltaSpec{} is a dynamic approach that combines test generation and specification inference, to automatically compute assertions affected by a commit change. In the following section we present the formal foundations regarding commit-relevant specifications and in Section~\ref{sec:computing-delta-specs} we describe the details of \DeltaSpec{}.


\section{Commit-Relevant Specifications}
\label{sec:delta-specs}


A \emph{commit-relevant specification} summarises the program properties that have changed as a
consequence of a specific software modification, via two sets of assertions, the \emph{delta-added} assertions and the \emph{delta-removed} assertions, that we formally define as follows.

\begin{defn}[Commit-Relevant Specification]
\label{def:commit-relevant-specification}
Let $P_{pre}$ and $P_{post}$ be the pre-commit and post-commit versions, respectively, of a target program $P$ modified by a commit change. Let $p$ be a program point common to $P_{pre}$ and $P_{post}$. Given a formal logical language ${\cal L}$, the \emph{commit-relevant specification} for $P_{pre}$ and $P_{post}$ at program location $p$, denoted by $\textit{DeltaSpec}_{p}^{\cal L}(P_{pre},P_{post})$, is defined as follows:

\vspace{-1.5em}
\begin{align*}
&\textit{DeltaSpec}_{p}^{\cal L}(P_{pre},P_{post}) =    \\
& \hspace*{1em}\textit{DeltaAdded}_{p}^{\cal L}(P_{pre},P_{post}) \cup \textit{DeltaRemoved}_{p}^{\cal L}(P_{pre},P_{post})\\
& \hspace*{.1em}\text{where: } \\
& \textit{DeltaAdded}_{p}^{\cal L}(P_{pre},P_{post})  =  \\
& \hspace*{8em} \{ \phi \in {\cal T}_{P_{post}, p}^{\cal L} | \phi \in {\cal S}_{P_{post}, p}^{\cal L} \land \phi \notin {\cal S}_{P_{pre}, p}^{\cal L} \}\\
&\textit{DeltaRemoved}_{p}^{\cal L}(P_{pre},P_{post}) = \\
& \hspace*{8em} \{ \phi \in {\cal T}_{P_{post}, p}^{\cal L} | \phi \notin {\cal S}_{P_{post}, p}^{\cal L} \land \phi \in {\cal S}_{P_{pre}, p}^{\cal L} \}
\end{align*}
\label{def-delta-spec}
\end{defn}
\vspace{-2.5em}
\noindent

Intuitively, $\textit{DeltaSpec}_{p}^{\cal L}(P_{pre},P_{post})$ aims at capturing how the set of assertions at program point $p$ evolve from the pre-commit to the post-commit version. $\textit{DeltaAdded}_{p}^{\cal L}(P_{pre},P_{post})$ characterises the set of properties that are valid in the post-commit version but were invalid in the pre-commit version. Added properties basically capture the new properties that emerge as a consequence of the just introduced changes. Conversely, $\textit{DeltaRemoved}_{p}^{\cal L}(P_{pre},P_{post})$ characterises the set of properties that were valid in the pre-commit but are invalid in the post-commit version. Removed properties capture the properties no longer valid as a consequence of the committed changes. Notice that the delta-added and delta-removed sets should be empty in case a commit is meant to be a code refactoring, i.e., when the committed changes are not meant to change the program behaviour. 

Besides the delta-added and delta-removed sets of assertions, we also define the set of valid properties that are not affected by the change: 

\begin{defn}[Preserved Specification]
\label{def:preserved-specification}
Let $P_{pre}$ and $P_{post}$ be the pre-commit and post-commit versions, respectively, of a target program $P$ modified by a commit change. Let $p$ be a program point common to $P_{pre}$ and $P_{post}$. Given a formal logical language ${\cal L}$, the \emph{preserved specification} at $p$ is defined as follows:
\begin{align*}
&\textit{Preserved}_{p}^{\cal L}(P_{pre},P_{post}) = \\
& \hspace*{8em} \{ \phi \in {\cal T}_{P_{post}, p}^{\cal L} | \phi \in {\cal S}_{P_{post}, p}^{\cal L} \land \phi \in {\cal S}_{P_{pre}, p}^{\cal L} \}
\end{align*}
\end{defn}

Finally, we define the task of automatically producing commit-relevant specifications for a given commit. 

\begin{defn}[Commit-aware Specification Inference]
\label{def:scommit-aware-spec-inference}
Let $P_{pre}$ and $P_{post}$ be the pre-commit and post-commit versions, respectively, of a target program $P$ modified by a commit change. Let $p$ be a program point common to $P_{pre}$ and $P_{post}$. Given a formal logical language ${\cal L}$, the \emph{commit-aware specification inference} problem for $P_{pre}$ and $P_{post}$ at location $p$ according to ${\cal L}$ is the task of automatically inferring assertions in the commit-relevant specification $\textit{DeltaSpec}_{p}^{\cal L}(P_{pre},P_{post})$.
\end{defn}

Notice that there is no specific assumption regarding the program point of interest for the inference problem. The definition applies to preconditions, postconditions, class invariants, as well as any other sort of program invariant, in the terminology of \cite{DBLP:journals/scp/ErnstPGMPTX07}.


\section{Inferring Commit-Relevant Specifications}
\label{sec:computing-delta-specs}

\begin{figure}[t]
\centering
\vspace{-0.5em}
\includegraphics[width=0.95\columnwidth]{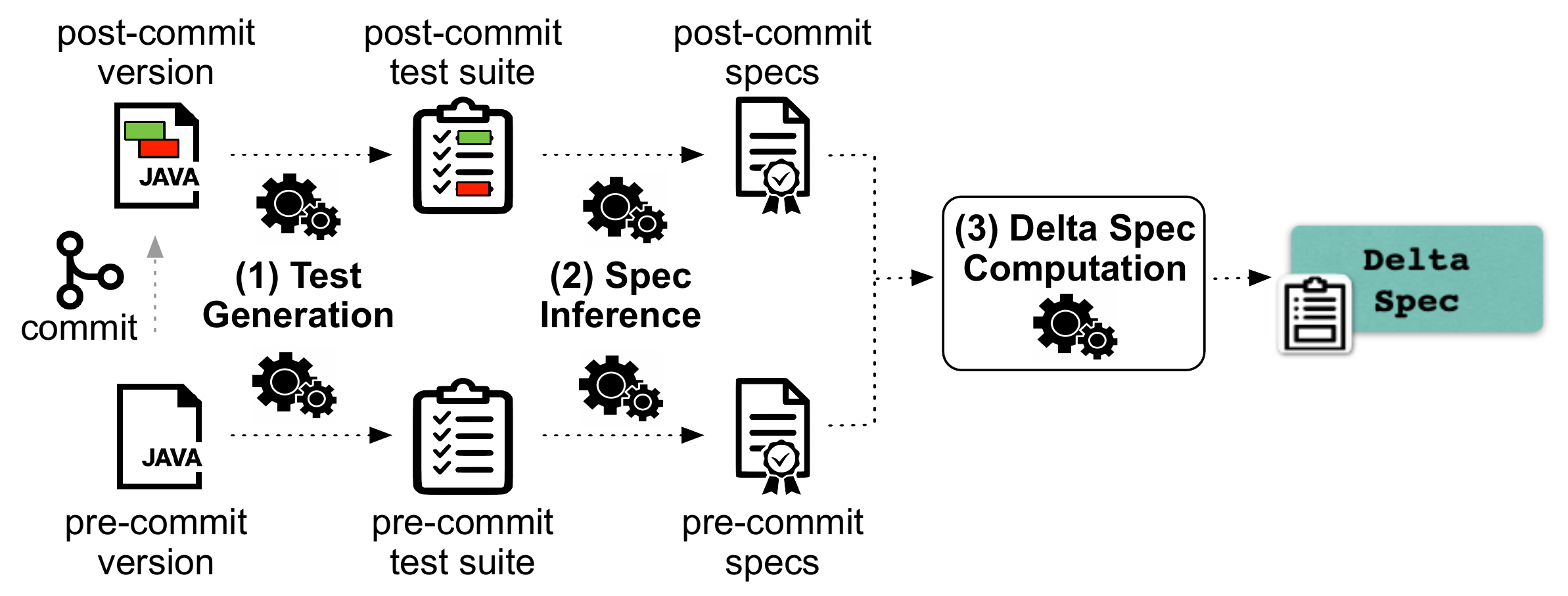}
\vspace{-0.8em}
\caption{Overview of \DeltaSpec's workflow.}
\label{fig:delta-spec-overview}
\vspace{-1.2em}
\end{figure}

Figure~\ref{fig:delta-spec-overview} shows an overview of \DeltaSpec's workflow. 
\DeltaSpec{} takes as input a target Java program $P$ and a target commit, from which we identify the two versions of the target program: the pre-commit version $P_{pre}$ and the post-commit version $P_{post}$. (Although the inference depends on a program location, we omit the reference to the location for simplicity.) The approach is organised as a pipeline of three steps: (i) a \emph{test generation} step that produces tests for both program versions, (ii) a \emph{specification inference} step that infers program specifications capturing the behavior of each version, and (iii) a semantic delta computation step that computes the commit-relevant specification for $P_{pre}$ and $P_{post}$. Below we discuss each one of these steps in  detail.

\subsection{Test Generation}
\label{sub-sec:test-gen-randoop}

The first step of \DeltaSpec{} consists of producing test suites exhibiting the behavior of each of the program versions. To do so, we first run  Randoop~\cite{Pacheco+2007}, a well-known automated test generation tool, to generate a test suite for 
the pre-commit version $P_{pre}$. Then, we repeat the same process and run Randoop for the post-commit version $P_{post}$. At the end of this step, we obtain two test suites $T_{pre}$ and $T_{post}$ for the $P_{pre}$ and $P_{post}$ program versions, respectively. Notice that the suites generated by Randoop are independent from the suites already present in the project under analysis. These newly produced suites are meant to thoroughly exercise the two program versions, and will be used for the purpose of inferring specifications for the two program versions. 
 
Although we use Randoop in our prototype, the user may replace it (or complement it) with other test generation tools, as well as with manually designed suites.

\subsection{Specification Inference}
\label{sub-sec:spec-inference-specfuzzer} 

\DeltaSpec{} performs specification inference to produce specifications capturing the behavior of the $P_{pre}$ and $P_{post}$ program versions. To do so, \DeltaSpec{} uses the SpecFuzzer specification inference tool~\cite{Molina+2022}. SpecFuzzer takes as input a target program and a test suite, and then infers specifications for the target program following three general steps. First, it extracts a grammar from the target program (constants, variable and function names, types, etc) and feeds it to a grammar-based fuzzer to automatically produce thousands of candidate assertions. Second, it passes the candidate assertions through a dynamic invariant detector in order to identify the ones that are consistent with the behaviour observed by the provided test suite. Finally, it uses mutation analysis in order to select the most relevant specifications by discarding redundant and irrelevant ones. At the end, SpecFuzzer outputs a set of assertions that are able to detect at least one mutant.

\DeltaSpec{} starts by considering a grammar $G$ for programs $P_{pre}$ and $P_{post}$, to ensure that the same \emph{language} of candidate assertions is used for both program versions. 
Then, \DeltaSpec{} feeds $G$ to SpecFuzzer's grammar-based fuzzer to generate a set $S$ of candidate assertions for programs $P_{pre}$ and $P_{post}$ (i.e., \DeltaSpec{}'s search space is complete with respect to to the finite set of generated assertions $S$). \DeltaSpec{} then runs SpecFuzzer taking as inputs the pre-commit program version $P_{pre}$, its corresponding suite $T_{pre}$, and the generated set of candidate assertions $S$, and returns the set of inferred assertions $S_{pre} \subseteq S$ that hold on program version $P_{pre}$. \DeltaSpec{} repeats the process and runs SpecFuzzer on inputs $P_{post}$, $T_{post}$ and $S$, inferring the set of assertions $S_{post} \subseteq S$  that hold on program version $P_{post}$. 

It is important to remark that since it is infeasible to generate every possible assertion for program $P_{post}$ (i.e., the assertion set $\mathcal{T}_{P_{post}}^{\cal L}$, which defines the assertion space for the commit-relevant specification), we \emph{approximate} it using the grammar-based fuzzer to produce the set $S$ of candidate assertions. 

\subsection{Commit-relevant Assertions Computation}
\label{sub-sec:compute-delta}  

This is the last and more important step of our approach. To compute a commit-relevant specification, \DeltaSpec{} uses the sets $S_{pre}$ and $S_{post}$ of inferred assertions for the $P_{pre}$ and $P_{post}$ program versions. According to Definition~\ref{def-delta-spec}, the commit-relevant specification is the union of the sets of added and removed assertions. Thus, in order to identify them, \DeltaSpec{} proceeds by computing the `diff' between the sets of assertions $S_{pre}$ and $S_{post}$ as follows. 

\emph{1) Added assertions} are those whose validity was introduced after the change, i.e., they are valid in the post-commit version $P_{post}$ but invalid in the pre-commit version $P_{pre}$. 
That is, $\textit{DeltaAdded}(P_{pre},P_{post})=S_{post}-S_{pre}$.

\emph{2) Removed assertions}  are those properties that were valid in the pre-commit version $P_{pre}$, but are no longer valid in the post-commit version $P_{post}$. 
That is, $\textit{DeltaRemoved}(P_{pre},P_{post})=S_{pre}-S_{post}$.

The union of the sets of added and removed assertions constitutes the commit-relevant specification for program versions  $P_{pre}$ and $P_{post}$. Notice that, in addition to the assertions that describe the semantic delta between program versions, we can also determine the set of assertions that were not affected by the commit:
$\textit{Preserved}(P_{pre},P_{post})=S_{pre} \cap S_{post}$. 

\subsection{Specification Language}
The commit-relevant specifications inferred by \DeltaSpec{} are limited by the language supported by SpecFuzzer~\cite{Molina+2022}. This language is similar in expressive power to the Java Modeling Language JML~\cite{DBLP:conf/fmco/ChalinKLP05}, and includes the usual relational, arithmetic and logical operators, as well as first-order quantification though the universal and existential quantifiers. The language's expressiveness is influenced by the expressive powers of the languages in other specification inference techniques, and in contract languages \cite{DBLP:journals/scp/ErnstPGMPTX07,gassert2020,Molina+2021,DBLP:conf/fmco/ChalinKLP05}.  

\DeltaSpec{}, all the scripts and data to reproduce the experiments discussed in the next sections, are publicly available in the accompanying replication package \cite{delta-spec-site}. 

\section{Research Questions}
\label{sec:RQs}

We start our experiments by analyzing the adequacy of commit-relevant specifications to describe code changes. Thus, we ask:  

\begin{description}
\item[RQ1] \emph{How adequate are commit-relevant specifications to explain semantic deltas of code modifications?}
\end{description}

To answer this question we discuss in detail several examples, for which we manually developed the corresponding commit-relevant specifications, and show how informative and precise they are to understand and validate the modifications. 

We continue our analysis by investigating the effectiveness of {\DeltaSpec} in inferring commit-relevant specifications. Thus, we ask: 

\begin{description}
\item[RQ2] \emph{How effective is {\DeltaSpec} in inferring commit-relevant specifications?}
\end{description}

To answer this question we start by running {\DeltaSpec} on the set of selected commits to infer the delta specifications. 
Then, we compare the inferred delta specifications with the set of \emph{manually written} specifications for RQ1, which are used as ground-truth (generated and cross-validated by the authors), and report the effectiveness of {\DeltaSpec} in resembling the same (or equivalent) specifications. 

We proceed to study an envisioned use case in which commit-relevant specifications are used for automatic analysis purposes. 
More precisely, we study the effectiveness of the inferred delta specs for testing the behaviours affected by the committed changes. That is, we study the ability of commit-relevant specifications to identify artificially seeded faults relevant to the commit changes, by using commit-aware mutation testing (cf. Section~\ref{sec:CI-testing}). 
Then, we ask:

\begin{description}
\item[RQ3] \emph{Are the inferred commit-relevant assertions more effective for detecting commit-relevant mutants than the specifications outside the delta?}
\end{description}

To answer this question we use as a ground-truth a publicly available dataset~\cite{DBLP:journals/tosem/OjdanicSDPL2022} that contains several commits and their respective commit-relevant mutants. 
Essentially, we perform a simulation where the tester analyses and selects assertions to use in a test suite, later run to assess the suite's mutation killing ability. Our goal is to measure the effectiveness of assertions for mutation killing, calculated in terms of the \emph{commit-relevant mutation score}, when the \emph{same number} of assertions are drawn from the pool of commit-relevant assertions, compared to the case in which these are taken from the pool of preserved assertions (i.e., those that are valid in the post-commit but are not part of the delta - cf.  Def.~\ref{def:preserved-specification}). 

Since part of the commit-relevant specification (the set of added assertions) is also valid in the post-commit version, it is natural to ask why not to use the entire post-commit specification for testing the change. Notice however that this may require more effort from the developer (more assertions to analyse), since the post-commit specification is a superset of the added assertions in the delta. Thus, we ask:
\begin{description}
\item[RQ4] \emph{How many assertions do we need to analyse to reach a given commit-relevant mutation score, when assertions are taken from the pool of added assertions, compared to doing so from the pool of all valid assertions of the post-commit?}
\end{description}

To answer this question we perform a simulation where we draw assertions to be analysed by developers. 
Essentially, we measure how many assertions are needed (the developer's cost) to reach the \emph{same rMS} (effectiveness), when assertions are taken from the pool of added assertions or the delta, compared to when they are taken from the entire set of valid assertions of the post-commit version.

\section{Experimental Setup}
\label{sec:exp-setup}


\subsection{Benchmarks and Ground Truth}
\label{subsec:subjects}

We consider the following projects from the Apache Commons~\cite{apache_commons}, a collection of projects of Java utility classes: commons-collections\footnote{https://github.com/apache/commons-collections}, commons-lang\footnote{https://github.com/apache/commons-lang} and commons-math\footnote{https://github.com/apache/commons-math}. 
These are mature open-source projects, include build infrastructure, with a large history of commits. 

To answer RQ1 and RQ2, we took some intuitive commits from well-known datasets, for which we were able to manually analyse and write the expected commit-relevant specifications. More precisely, we took some of the bugs reported in Defects4J \cite{defects4j}, one of the largest collections of reproducible real faults for Java programs, for commons-lang and commons-math projects. 
We selected 3 faults from commons-lang, containing 2 classes and 7 methods, and 9 faults from commons-math, with 10 classes and 14 methods. 
We also took 6 commits from commons-collections, with 6 classes and 6 methods, involving some refactoring commits, where the expected commit-relevant specification for these commits should be empty (i.e., no added or removed properties should arise). 
Table~\ref{tab:d4j} summarises the number of commits, classes and methods we analysed in RQ1 and RQ2.

To answer RQ3 and RQ4, we used the publicly available dataset\footnote{https://mutationtesting-user.github.io/evolve-mutation.github.io/} provided by Ojdanic et al.~\cite{DBLP:journals/tosem/OjdanicSDPL2022}. It includes a list of commits, the set of modified classes by the commit, and the labels for each mutant generated with PiTest~\cite{pitest} that indicates if it is relevant or not for the commit. 
We use mutants' label to assess the effectiveness of \DeltaSpec{} in testing the change, that is, the ability of the inferred commit-relevant specifications for killing commit-relevant mutants. 
Table~\ref{tab:commit-relevant-mutants} summarises the number of commits, classes and  methods we consider in our analysis.

\begin{table}[t]
\centering
\caption{Commits where we manually derived the commit-relevant specifications (RQ1). Used as ground-truth in RQ2.}
\resizebox{\columnwidth}{!}{
\begin{tabular}{lrrr}
\toprule
\textbf{Project} & \textbf{Commits} & \textbf{Classes} & \textbf{Methods} \\
\midrule
commons-collections & 6 & 6 & 6 \\
commons-lang & 3 & 2 & 7  \\
commons-math & 9 & 10 & 14  \\
\midrule
\textbf{Total} & \textbf{18} & \textbf{18} & \textbf{27} \\
\bottomrule
\end{tabular}}
\label{tab:d4j}
\end{table}

\begin{table}[t]
\centering
\caption{Dataset used to simulate a commit-aware mutation testing scenario in RQ3 and RQ4.}
\resizebox{\columnwidth}{!}{
\begin{tabular}{lrrrrr}
\toprule
\textbf{Project} & \textbf{Commits} &  \textbf{Classes} & \textbf{Methods} \\
\midrule
commons-collections & 15  &  19  & 142 \\
commons-lang        & 24  & 26  & 628  \\
\midrule
\textbf{Total} & \textbf{39} & \textbf{45} &  \textbf{770} \\
\bottomrule
\end{tabular}}
\label{tab:commit-relevant-mutants}
\end{table}

\subsection{Experimental Procedure}
\label{subsec:experimental-procedure}



To answer RQ1, we inspect each one of the selected 18 subjects (Table~\ref{tab:d4j}) and then manually elaborate a commit-relevant specification that can be used for validating the commit change. 
To do so, we basically focus on every information provided by the commit, such as the commit message, the extra test cases added to the suite, and of course the commit code changes in order to write the specification. 
Later on we show some of the manually developed specifications and we provide the full list in the accompanying replication package.

The experiment for RQ2 consists in running \DeltaSpec{} on each selected subject, and analysing the percentage of manually written specifications (ground truth) that the tool is able to infer. 
We manually inspect the commit-relevant specifications generated by \DeltaSpec{} and verify if these are present in the ground truth (or are equivalent to some of them) and, conversely, we also check if all ground truth specifications are captured by the \DeltaSpec{}'s inferred assertions. The comparison is not merely syntactic; we use SMT Solving to check if the same assertion, or one equivalent, appears in the ground truth (and vice versa).


To answer RQ3 and RQ4, we start by running \DeltaSpec{} on the classes modified by the commits taken from the dataset published in the work of Ojdanic et al.~\cite{DBLP:journals/tosem/OjdanicSDPL2022} (see Table~\ref{tab:commit-relevant-mutants}). 
Given $P_{pre}$ and $P_{post}$, the pre-commit and post-commit versions of a program $P$, \DeltaSpec{} will compute the three set of assertions $DeltaAdded(P_{pre},P_{post})$, $DeltaRemoved(P_{pre},P_{post})$ and $Preserved(P_{pre},P_{post})$, as explained in Section~\ref{sec:delta-specs}. 
Then, we proceed to generate and run the mutants for the modified classes on the post-commit version, to check what are the mutants killed by each inferred assertion. 
Recall that an assertion kills a mutant if the mutant's execution violates the assertion (i.e., the assertion is not valid on the mutant). In the same way as in mutation testing, where it is required that the original program passes the test suite, we only use the assertions that are valid on the post-commit version. That is, we check which mutants are killed by each assertion from $DeltaAdded(P_{pre},P_{post})$ and $Preserved(P_{pre},P_{post})$ sets, but we do not consider assertions from $DeltaRemoved(P_{pre},P_{post})$, since these are by definition invalid in the post-commit version. 
 

Finally, we perform a controlled experiment that simulates a scenario where a tester takes assertions from the pool of inferred assertions, includes them into the test suite and runs mutation testing to determine which mutants are killed. 

When answering RQ3, we control the \emph{number of selected assertions}. 
That is, we simulate a scenario where the tester selects the same number of assertions from the pool of added and preserved assertion that are used for analysis. Then, we compute the commit-relevant mutation score (rMS) for each set of selected assertions, to study if delta-added assertions or preserved assertions are more effective for testing the change. 
We consider different selection sizes (from 1 to 10) that are reasonable for manual analysis, and repeat the experiment 100 times (for each selection size) to avoid coincidental results. 

When answering RQ4, we control the rMS to achieve.  
That is, we simulate a scenario where the tester selects assertions to kill mutants until it achieves a given rMS. 
Basically, we study and compare the number of assertions to select from the set of delta-added assertions and the entire set of valid assertions of the post-commits, in order to reach same effectiveness (same rMS). 
Our goal is to study what is the overhead when the testing process is guided by the entire pool of valid assertions of the post-commit, and not by the commit-relevant specifications.


To study the correlation between the rMS obtained by different sets of assertions, we use the Mann-Whitney U-test~\cite{MannWhitneyUTest} and Vargha-Delaney A measure $\hat{A}_{12}$~\cite{VarghaDelaney2000}. 

\section{Results}
\label{sec:results}

\subsection{Commit-relevant specs for commit validation (RQ1)}

Out of 18 analysed cases, 4 of them were refactorings, and 14 were cases in which the program behaviour was altered. Let us first analyse the refactorings. In such cases, as it is expected, the program behaviour is not changed, and thus a commit-aware specification explaining such changes should be empty. For instance, consider commit \CodeIn{fc3d530}\footnote{https://github.com/apache/commons-collections/commit/fc3d530} shown in Figure~\ref{example-commons-collections-6-delta}, from the commons-collection project. The change is modifying the \CodeIn{remove} method of class \CodeIn{SingletonListIterator}, by removing an else clause and moving the statements that were inside the else to the end of the method. For this case, it is easy to see that the delta specification is empty, indicating that the commit is not changing the program behaviour. However, for other refactorings, especially those involving changes in many classes and methods, ensuring that the delta is empty can be more challenging. 

\begin{figure}[tp]
\vspace{-0.5em}
\centering
\begin{subfigure}{.2\textwidth}
\begin{lstlisting}[language=diff]
public void remove() {
  if (!nextCalled || remove) {
    throw new IllegalStateException();
- } else {
-   object = null;
-   removed = true;
  }
+ object = null;
+ removed = true;
}
\end{lstlisting}
\vspace{-1.0em}
\caption{\texttt{remove} method.}
\label{changed-method-remove}
\end{subfigure}
\begin{subfigure}{.2\textwidth}
\begin{lstlisting}[language=diff,escapeinside={(*}{*)},numbers=none]
            (*$\emptyset$*)
\end{lstlisting}
\vspace{-1.0em}
\caption{Commit-relevant spec.}
\label{delta-spec}
\end{subfigure}
\vspace{-0.5em}
\caption{Commit \CodeIn{fc3d530} and its commit-relevant specs.}
\vspace{-1.0em}
\label{example-commons-collections-6-delta}
\end{figure}

Let us now consider commit \CodeIn{38140a5}\footnote{https://github.com/apache/commons-lang/commit/38140a5}, which fixes a bug\footnote{https://issues.apache.org/jira/browse/LANG-1271} in the commons-lang project. 
A fragment of this commit is shown in Figure~\ref{example-commons-lang-d4j-182-delta}(a), indicating a specific modification in method \CodeIn{isAnyEmpty}. The bug is related to the fact that method \CodeIn{isAnyEmpty} is returning the wrong boolean value when it is invoked with an empty array. 
Intuitively, \CodeIn{isAnyEmpty} is an existential quantifier (``any'') that should return \CodeIn{false} (the neutral value for the disjunction) when there is no element in the array.  
Figure~\ref{example-commons-lang-d4j-182-delta}(b) shows the commit-relevant specification we manually developed for such commit. We see that the specification is precisely capturing the change, and that it is consistent with the described issue. 

\begin{figure}[tp]
\vspace{-0.5em}
\centering
\begin{subfigure}{.4\textwidth}
\begin{lstlisting}[language=diff]
public static boolean isAnyEmpty(final CharSequence... css) {
  if (ArrayUtils.isEmpty(css)) {
-   return true; 
+   return false;
  }
  for (final CharSequence cs : css){
    if (isEmpty(cs)) {
    ...
\end{lstlisting}
\vspace{-1.0em}
\caption{\texttt{isAnyEmpty} method.}
\label{changed-method-isanyempty}
\end{subfigure}
\begin{subfigure}{.4\textwidth}
\begin{lstlisting}[language=diff]
- ArrayUtils.isEmpty(css) implies result
+ ArrayUtils.isEmpty(css) implies !result
\end{lstlisting}
\vspace{-1.0em}
\caption{Commit-relevant spec.}
\label{delta-spec}
\end{subfigure}
\vspace{-0.5em}
\caption{Commit \CodeIn{38140a5} and its commit-relevant specs.}
\vspace{-0.5em}
\label{example-commons-lang-d4j-182-delta}
\end{figure}

In other cases, identifying the commit-relevant specification for the whole commit was more challenging. For instance, consider the fragment of commit \CodeIn{a21d5ae}\footnote{https://github.com/apache/commons-math/commit/a21d5ae} 
illustrated in Figure~\ref{example-commons-math-d4j-327-delta}(a). This commit was performed to fix a bug\footnote{https://issues.apache.org/jira/browse/MATH-692} in commons-math, related to the computation of cumulative and inverse cumulative probabilities. The shown fragment fixes the implementation of method \CodeIn{getSupportUpperBound}, used by the methods that compute the mentioned probabilities. 
Although it was not easy to come up with the exact specification, the delta assertions describing the semantic change on method \CodeIn{getSupportUpperBound} are shown in Figure~\ref{example-commons-math-d4j-327-delta}(b).    

\begin{figure}[tp]
\vspace{-0.5em}
\centering
\begin{subfigure}{.4\textwidth}
\begin{lstlisting}[language=diff]
  public int getSupportUpperBound() {
-   return getNumberOfTrials();
+   return probabilityOfSuccess > 0.0 ? numberOfTrials : 0;
  }
\end{lstlisting}
\vspace{-1.0em}
\caption{\texttt{getSupportUpperBound()} method.}
\label{changed-method-getsupport}
\end{subfigure}
\begin{subfigure}{.4\textwidth}
\begin{lstlisting}[language=diff]
- result = numberOfTrials
+ probabilityOfSuccess > 0 implies result = numberOfTrials
+ probabilityOfSuccess <= 0 implies result = 0
\end{lstlisting}
\vspace{-1.0em}
\caption{Commit-relevant spec.}
\label{delta-spec}
\end{subfigure}
\vspace{-0.5em}
\caption{Commit \CodeIn{a21d5ae} and it's commit-relevant specs.}
\label{example-commons-math-d4j-327-delta}
\vspace{-0.5em}
\end{figure}

In these examples we illustrated how commit-relevant specifications can be used to properly capture the impact that a change is having on the code base being affected, providing a useful mechanism for code change validation. 

\begin{tcolorbox}[standard jigsaw, opacityback=0]
Answer to RQ1: Commit-relevant specifications can precisely describe the semantic deltas produced by code modifications, providing a useful mechanism for code validation.
\end{tcolorbox}

\subsection{\DeltaSpec{}'s inference effectiveness (RQ2)}
 
For the 18 commits we analysed (Table~\ref{tab:d4j}), we produced a total of 56 ground truth properties, with 29 added and 27 removed. Only 17 out of the 56 were expressible in the specification language of our technique. The remaining 39 were properties whose specifications would require array indexing, type determination over variables and exception handling; all characteristics that cannot be expressed in our current specification language. 
Table~\ref{tab:delta-spec-effectiveness} summarises the results of running \DeltaSpec{} on the selected subjects. 
\DeltaSpec{} is able to infer 15 out of the 17 manually written delta assertions that are expressible in the language used by the tool. 
This presents an effectiveness of 88\% with respect to the 17 expressible assertions, and the 27\% of the entire set of manually developed assertions previously use in RQ1. 

In our replication package we provide the full list of manually written assertions, indicating which of them can and cannot be expressed in the \DeltaSpec{}'s language, as well as the commit-relevant specifications inferred by the tool for each of the target commits.

\begin{table}[t]
\centering
\caption{Effectiveness of \DeltaSpec{}.}
\resizebox{\columnwidth}{!}{
\begin{tabular}{lrrrr}
\toprule
\multirow{2}{*}{\textbf{Project}} &
\multicolumn{2}{c}{\textbf{Ground Truth}} &
\multicolumn{2}{c}{\textbf{\DeltaSpec{}}} \\
& \textbf{\# All} & \textbf{\# Expressible} & \textbf{Inferred (\%)}  & \textbf{Expressible Inferred (\%)} \\
\midrule
commons-collections & 2 & 2 & 100\% & 100\%  \\
commons-lang & 18 & 8 & 44\% & 100\% \\
commons-math &  36 & 7 & 13\% & 71\% \\
\midrule
\textbf{Total} & 56 & 17 & 27\% & 88\% \\
\bottomrule
\end{tabular}
}
\label{tab:delta-spec-effectiveness}
\vspace{-1.0em}
\end{table}

\begin{tcolorbox}[standard jigsaw, opacityback=0]
Answer to RQ2: \DeltaSpec{} infers 88\% of the manually written delta specifications, that can be expressed in \DeltaSpec{}'s supported specification language. 
\end{tcolorbox}

\subsection{Comparing the rMS obtained when (same number of) assertions are selected from inside and outside the commit-relevant specification (RQ3)}

Table~\ref{tab:inferred_specs_summary} summarises the number of analysed cases and the number of non-empty inferred commit-relevant specifications (between brackets) for which we perform the mutation analysis. In total, for 22 out of 39 commits \DeltaSpec{} produces a non-empty delta specification, including 24 classes (out of the 45) and 129 methods (out of the 770), for which we analyse a total of 17212 mutants, 3314 of which are commit-relevant.

\begin{table}[t]
\centering
\caption{\DeltaSpec{} infers a non-empty commit-relevant specification for 22 out of the 39 analysed commits.}
\resizebox{\columnwidth}{!}{
\begin{tabular}{lrrrrr}
\toprule
\textbf{Project} & \textbf{Commits} &  \textbf{Classes} & \textbf{Methods} &  \textbf{Mutants} & \textbf{Relevant}\\
\midrule
commons-collections & 15 (9) &  19 (9) & 142 (32) & 1494 & 649\\
commons-lang        & 24 (13) & 26 (15) & 628 (97) & 15718 & 2665 \\
\midrule
\textbf{Total} & \textbf{39 (22)} & \textbf{45 (24)} &  \textbf{770 (129)} & \textbf{17212} & \textbf{3314}\\
\bottomrule
\end{tabular}
}
\label{tab:inferred_specs_summary}
\vspace{-1.0em}
\end{table}

\begin{figure}[htp!]
\vspace{-0.6em}
\centering
\includegraphics[width=0.8\columnwidth]{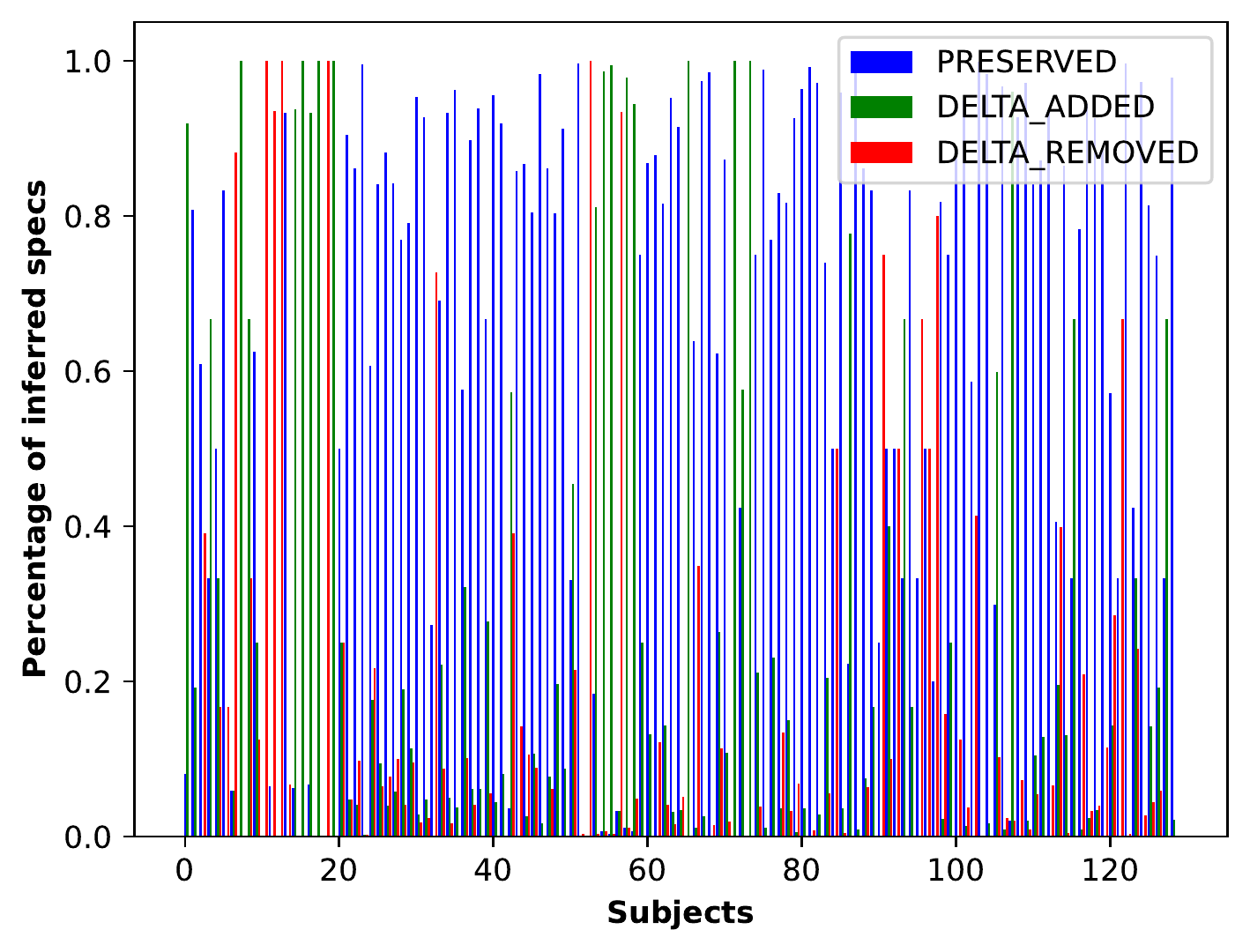}
\vspace{-1.0em}
\caption{Percentage of added, removed and preserved assertions inferred by \DeltaSpec{}.}
\label{fig:inferred_specs_summary}
\vspace{-0.7em}
\end{figure}

Figure~\ref{fig:inferred_specs_summary} summarises the percentage of delta-added, delta-removed and preserved inferred assertions with respect to the total number of inferred assertions for each subject. 
On average, 23\% of the inferred assertions are delta-added properties, 15\% of the assertions are delta-removed properties, while the remaining 62\% of the assertions are preserved properties (outside the delta, valid in both the pre and post-commit). 

\begin{figure}[htp!]
\vspace{-0.6em}
\centering
\includegraphics[width=0.8\columnwidth]{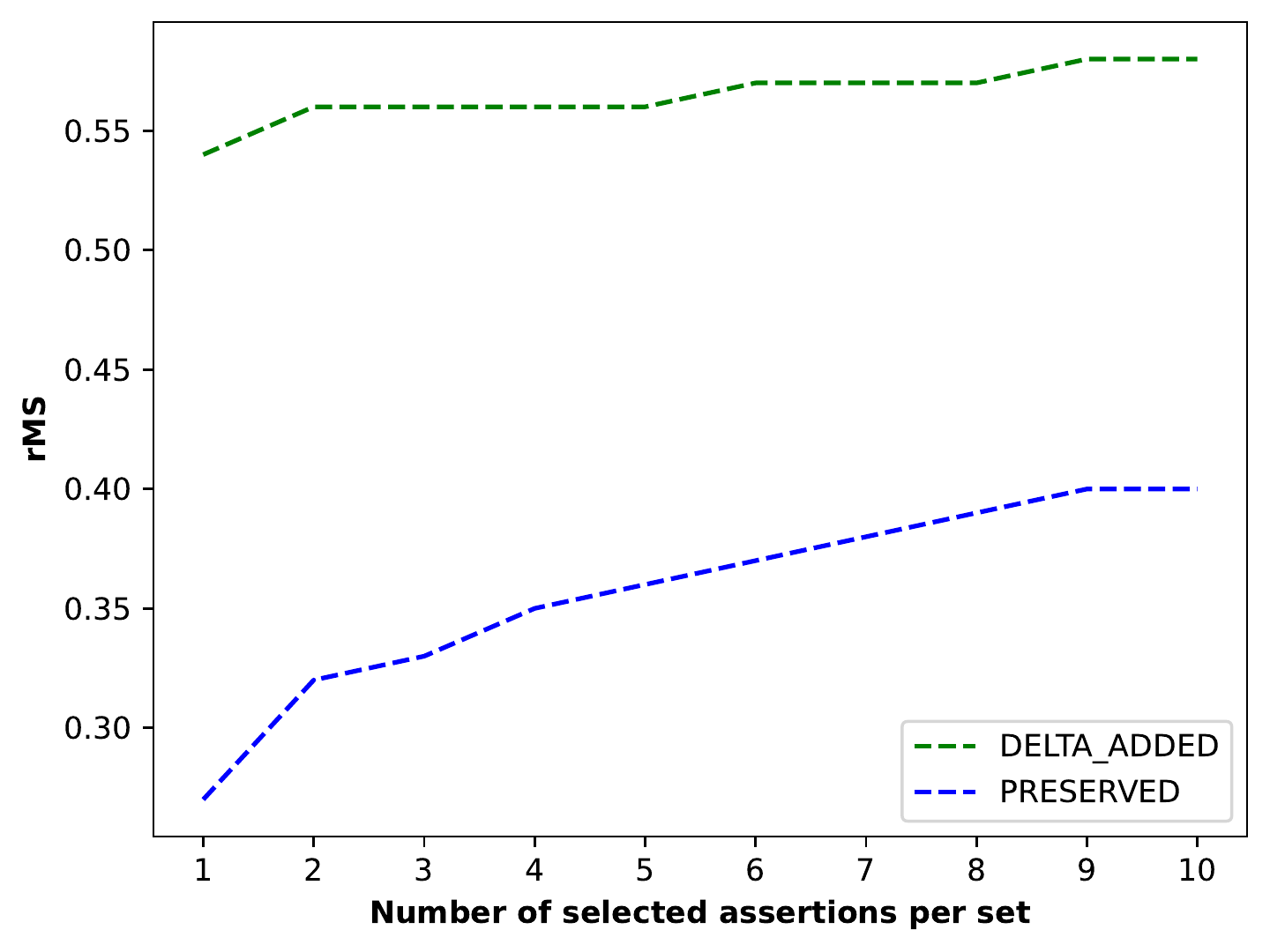}
\vspace{-1.0em}
\caption{Relevant-mutation score (rMS) obtained by \DeltaSpec{}'s  inferred  specifications when same number of assertions are selected.}
\label{fig:rMS_summary}
\vspace{-0.5em}
\end{figure}

Figure~\ref{fig:rMS_summary} summarises the average rMS obtained by the selected delta-added and preserved assertions during the simulation performed to answer RQ3. 
We can observe that when only one assertion is selected, the delta-added specification obtains, on average, 54\% of rMS while the preserved one obtains 27\%.
That is, the rMS obtained by the delta specification is, on average, 100\% higher when only 1 assertion is selected. 
When 5 assertions are selected, the delta specification obtains, on average, 56\% of rMS while the preserved specification obtains 36\%, i.e., a 55\% of improvement. 
While, when 10 assertions are selected, the improvements is of 48\% in the rMS.
The differences between the rMS obtained by the sets of assertions are statistically significant. 

These results suggest that assertions from the inferred commit-relevant specification are more likely to find faults that interact with the commit change, than assertions that are not part of the delta.

\begin{tcolorbox}[standard jigsaw, opacityback=0]
Answer to RQ3: Commit-relevant specifications are, on average, 58.3\% more effective in finding commit-relevant mutants than specifications that are preserved by the change.
\end{tcolorbox}

\subsection{Comparing the number of selected assertions in order to reach same rMS (RQ4)}
\label{sec:effort-comparison}

\begin{figure}[htp!]
\vspace{-0.6em}
\centering
\includegraphics[width=0.8\columnwidth]{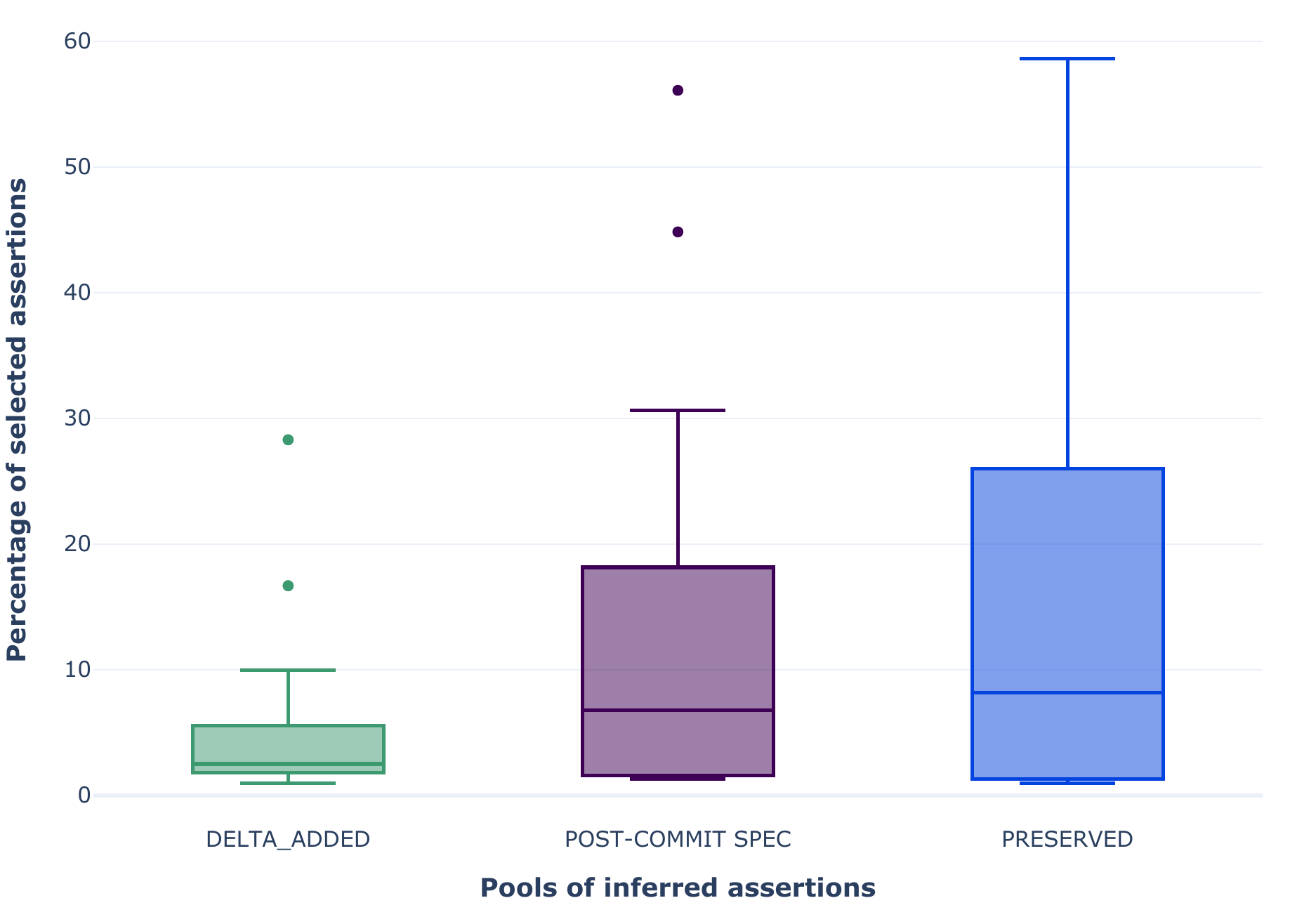}
\vspace{-1.0em}
\caption{Percentage of selected assertions to reach same rMS.}
\label{fig:effort_comparison}
\vspace{-1.0em}
\end{figure}

Figure~\ref{fig:effort_comparison} summarises the number of assertions selected from each pool, to achieve the same rMS. 
On average, to reach the same rMS we need to select 2.52, 6.78, and 8.2 assertions from the pools of delta-added assertions, all valid assertions in the post-commit, and preserved assertions, respectively. This suggests that, if we guide the testing process by the commit-relevant assertions, the number of selected assertions is reduced by 169\% and 225\%, compared to when we pick assertions from all the post-commit valid assertions, and preserved assertions, respectively, without losing effectiveness. 

\begin{tcolorbox}[standard jigsaw, opacityback=0]
Answer to RQ4: Selecting commit-relevant assertions can help in reducing by 169\% and 225\% the effort, without reducing rMS, compared to selecting assertions from the 
pool of valid assertions of the post-commit and preserved assertions, respectively.
\end{tcolorbox}

\section{Threats to Validity}
\label{sec:threats-to-validity}

Threats to \emph{external validity} may arise from the subjects we used and the generalisability of the results to other programs and languages. 
The subject commits we took from Defects4J, analysed in RQ1 and RQ2, were selected among others because they were simple and intuitive changes that allowed us to manually write commit-relevant specifications (later on used in the evaluation). 
When selecting commits from the dataset containing the information of commit-relevant mutants~\cite{DBLP:journals/tosem/OjdanicSDPL2022}, we do not consider commits from commons-io, commons-text and commons-csv, due to SpecFuzzer's grammar~\cite{Molina+2022} limitations, which cannot produce assertions for strings and classes handling files. Incorporating them would require additional work on SpecFuzzer's grammar. 

\emph{Internal validity} threats may relate to the stochastic nature involved in parts of \DeltaSpec{} implementation (SpecFuzzer~\cite{Molina+2022} and Randoop~\cite{DBLP:journals/scp/ErnstPGMPTX07}). To mitigate this threat we make publicly available our implementation, repeated several times the experiments, and manually validated our results. 

\textit{Construct Validity} threats may relate to the manually written commit-relevant specifications we developed (RQ1 and QR2). To mitigate this threat, we made the specifications publicly available in our replication package. 
Other construct validity threats may relate to our assessment metrics, such as the number of analysed assertions and the commit-relevant mutation score, may not reflect the actual testing cost/effectiveness values. 
However, these metrics have been widely used in the literature \cite{PapadakisK00TH19,AndrewsBLN06,KurtzAODKG16} and are intuitive, since the number of analysed assertions essentially simulates the manual effort put by a developer, while the test suites developed to kill mutants can also be used to measure its effectiveness in finding faults that are relevant to the change. 
In our experiments, test cases were automatically generated by Randoop~\cite{DBLP:journals/scp/ErnstPGMPTX07}, which may not reflect the real cost/effort in designing such test cases.

\section{Related Work}
\label{sec:related-work}


Specification inference is an active area of research. Besides the techniques that infer contract assertions, such as SpecFuzzer~\cite{Molina+2022},  Daikon~\cite{DBLP:journals/scp/ErnstPGMPTX07}, Jdoctor~\cite{Blassi+2018}, GAssert~\cite{gassert2020} and EvoSpex~\cite{Molina+2021}), there are also other approaches focusing on generating other kinds of specifications, or from other sources. For instance, some approaches monitor software behaviour and attempt to infer test oracles~\cite{FraserZ12, DBLP:conf/icse/WatsonTMBP20} (that is, assertions that are only valid in specific unit tests), while others rely on modern machine learning models to statically generate context-dependent test oracles~\cite{DBLP:conf/icse/DinellaRML22}. 
Other works are instead focused on producing abstractions of the software behaviour that can be used for validation and test generation~\cite{DBLP:journals/tosem/CasoBGU13,Lo2018,Lo2021}. 
Other techniques attempt to infer weaker oracles in the form of metamorphic properties, from code comments instead of from source code~\cite{BlasiGEPC2021, Blassi+2018}.  
All these techniques could complement the specification inference process within \DeltaSpec{} in order to produce other types of delta specifications. 

Various studies proposed to use coverage instead of mutation, to analyse the impact on control and data dependencies affected by changed code~\cite{RothermelH94,Binkley97}. 
When programs evolve frequently and new features are included, test augmentation approaches aim at generating new test cases that, for instance, trigger an unseen program output~\cite{QiRL10}, increase coverage~\cite{XuKKRC10} or increase  mutation score~\cite{SmithW09JSS,SmithW09EMSE}.

Commit-aware mutation testing aims at defining commit-relevant mutants, i.e., mutants affected by a commit. 
While Petrovic et al.~\cite{petrovic2018} proposed to use only the mutants located on the changed lines, recent works have shown that a big proportion of mutants interacting with the change are outside the changed lines~\cite{DBLP:journals/ese/OjdanicMLCVP22,DBLP:journals/tosem/OjdanicSDPL2022}. 
Precisely, our definition of commit-relevant assertions is in line with the definition of commit-relevant mutants proposed by Ma et al.~\cite{ma2020commit} that considers that a mutant is relevant to a commit if its behaviour is different from the pre-commit to the post-commit version. Recently, a machine learning approach was presented to predict whether a mutants is or not a commit-relevant mutant~\cite{ma2021mudelta}. 
Our results show that \DeltaSpec{}'s inferred assertions are effective in killing commit-relevant mutants, and may help in identifying them. However, studying the relation between delta assertions and commit-relevant mutants is outside the scope of this paper.

\section{Conclusion}
\label{sec:conclusion}
We introduced the notion of \emph{commit-relevant specifications} as a mean to explain the semantic delta introduced by commits/changes. We also presented \DeltaSpec{}, a dynamic approach that combines test generation and specification inference to infer commit-relevant assertions. We showed that \DeltaSpec{} can be effective in producing manually written specifications (88\% of the assertions expressible in the supported language), and can be applied on large programs (e.g., the subjects taken from the commons family) to equip them with commit-relevant specifications.  When used for testing, they can reach on average the 78.3\% of rMS. Moreover, delta assertions can reach, on average, 58.3\% higher rMS than assertions outside the delta; and in order to reach a same effectiveness (same rMS), 126\% less delta assertions are required, compared to when these are randomly picked from the pool of valid assertions on the post-commit version. 

As part of future work, we plan to extend the grammar used by \DeltaSpec{} to produce a richer set of properties. We also plan to study the relationship between the delta assertions and the commit-relevant mutants.

\bibliographystyle{plain}
\bibliography{bibfile}
\balance
\end{document}